\documentclass[preprintnumbers,amsmath,amssymb,floatfix,11pt,prd,onecolumn,
superscriptaddress,nofootinbib]{revtex4}
\usepackage{graphicx}
\usepackage{epsfig}
\usepackage{bm}
\usepackage{amsfonts}

\begin{document}

\title{\textbf{ Statefinder Analysis of f(T) Cosmology}}

\author{\textbf{ Mubasher Jamil}} \email{mjamil@camp.nust.edu.pk}
\affiliation{Center for Advanced Mathematics and Physics (CAMP),\\
National University of Sciences and Technology (NUST), H-12,
Islamabad, Pakistan}
\affiliation{Eurasian International Center
for Theoretical Physics, Eurasian National University, Astana
010008, Kazakhstan}

\author{\textbf{ Davood Momeni}}
\email{d.momeni@yahoo.com}
 \affiliation{Eurasian International Center
for Theoretical Physics, Eurasian National University, Astana
010008, Kazakhstan}

\author{\textbf{ Ratbay Myrzakulov}}
\email{ rmyrzakulov@csufresno.edu}\affiliation{Eurasian
International Center for Theoretical Physics, Eurasian National
University, Astana 010008, Kazakhstan}

\author{ \textbf{Prabir
Rudra}}\email{prudra.math@gmail.com} \affiliation{Department of
Mathematics, Bengal Engineering and Science University, Shibpur,
Howrah-711 103, India.}

\begin{abstract}

\textbf{Abstract:} In this paper, we intend to evaluate and analyze
the statefinder parameters in $f(T)$ cosmology.  Friedmann equation
in  $f(T)$ model is taken, and the statefinder parameters $\{r,s\}$
are calculated. We consider a model of $f(T)$ which contains a
constant, linear and non-linear form of torsion. We plot $r$ and $s$
in order to characterize this model in the $\{r,s\}$ plane. We found
that our model $f(T)=2C_1 \sqrt{-T} +\alpha T+C_2,$ predicts the
decay of dark energy in the far future while its special case namely
teleparallel gravity predicts that dark energy will overcome over
all the energy content of the Universe.

\end{abstract}

\pacs{04.20.Fy; 04.50.+h; 98.80.-k}
\maketitle

\newpage

\section{Introduction}

The incompatibility of General Relativity (GR) as a complete
theory of gravity came to light when the recent cosmological
observations from Ia supernovae, CMBR via WMAP, galaxy redshift
surveys via SDSS and and galactic X-ray indicated that the
observable Universe enters into an epoch of accelerated expansion
\cite{c1, c2, c3, c4}. In the quest of finding a suitable model
for Universe, cosmologists started to investigate the root cause
that is responsible for this expansion. Existence of a mysterious
negative pressure component which violates the strong energy
condition i.e. $\rho+3p<0$ is proposed as one of the major
alternatives responsible for the recent cosmic acceleration.
Because of its invisible nature this energy component is aptly
termed as dark energy (DE) \cite{Riess1} (also see a recent review
on DE \cite{bamba1}). Quite surprisingly, observations show that
about 70 percent of the Universe is filled by this unknown
ingredient and in addition that about 25 percent of this is
composed by dark matter (DM).

The simplest model for the accelerating universe is the Universe
associated with a tiny cosmological constant $\Lambda$ \cite{c7}.
But it was soon found that the model suffered from fine tuning and
cosmic coincidence problems. So the search for a better model
continued. With the passage of time, various models of DE were
proposed and subsequently developed. Some of them are quintessence
scalar fields \cite{quint}, Chaplygin gas \cite{chaplygin},
phantom energy field \cite{phant}, f-essence \cite{f} and
interacting dark energy models \cite{obs, cr, pr}. As so many
cosmological DE models started appearing in the scene gradually,
it became an utmost necessity to devise a method that will be able
to either qualitatively or quantitatively differentiate between
the various DE models. Hence in order to bring about this
discrimination between the various contenders, Sahni et al
\cite{Sahni1} proposed a new geometrical diagnostic named the
statefinder pair $\left\{r, s\right\}$. Clear differences for the
evolutionary trajectories in the $r-s$ plane have been found for
different dark energy models in \cite{jamil}. The statefinder
parameters are defined as follows,
\begin{equation}\label{state8.1}
r\equiv\frac{\stackrel{...}a}{aH^3}~~~~~~~~~~s\equiv\frac{r-1}{3\left(q-\frac{1}{2}\right)},
\end{equation}
where $a$ is the scale factor of the Universe, $H$ is the Hubble
parameter, a dot denotes differentiation with respect to the cosmic
time $t$ and $q$ is the deceleration parameter given by,
\begin{equation}\label{state8.2}
q=-\frac{\dddot a}{aH^{2}}.
\end{equation}
It is found that for the $\Lambda$CDM model, $r = 1$ and $s = 0$.

In order to bring about the discrimination between various DE
models the trajectories in the $r - s$ plane for these models are
generated. The distances of the contrasting $r-s$ trajectories of
these models from $(0,1)$ point is computed. These computed
distances, as expected, vary significantly from model to model,
thus giving us an extremely useful tool to differentiate between
various cosmological DE models. Thus the difference in the
trajectory from the standard $\Lambda$CDM model produces the
required discrimination and characterizes the given model up to a
great extent \cite{pano}.

A section of cosmologists concentrated on the left hand side of
the Einstein's equation ($G_{\mu\nu}=8\pi GT_{\mu\nu}$) rather
than the right hand side in order find an effective modification
that will admit the recent cosmic acceleration in its framework.
They ended up modifying the Einstein gravity and giving birth to
various modified gravity theories. This alternative method along
with the concept of DE have gained enormous popularity in the
cosmological society since it passes several solar system and
astrophysical tests successfully \cite{sergei2}. It is found that
both the methods (DE and modified gravity models) can successfully
account for the recent cosmic acceleration independently. In the
context of modified gravity theory it is worth stating that Loop
quantum gravity \cite{lqc}, extra dimensional braneworlds
\cite{brane}, $f(R)$ \cite{fr}, $f(T)$ \cite{ft, miao, miaoli,
sotirio} are some of the popular modified gravity models.

An attempt to unify gravitation with electromagnetism gave birth
to Teleparallelism (Einstein). But as the years passed by
scientists lost interest in this concept of unification because of
the conceptual and physical diversity of various physical
theories. As a result, today teleparallelism stands just as a
theory of gravity. Teleparallel gravity corresponds to a gauge
theory for the translation group \cite{trans}. Einstein introduced
the crucial new idea of a tetrad field in this theory. Here the
space-time is characterized by a curvature-free linear connection,
called the Weitzenbock connection. Since the space-time is free
from curvature, it is quite evident that the theory in its
physical aspect is quite different from the Einstein gravity,
whose main feature is curvature. In Teleparallelism, curvature is
in fact substituted by torsion, another physical tool responsible
for the dynamics of space-time. In the framework of general
relativity (GR), curvature is used to geometrize space-time, thus
successfully describing the gravitational interaction between
particles. But in Teleparallelism, the role of gravitation is
played by torsion not by geometrizing the interaction (unlike
curvature), but by merely acting as a force. This implies that, in
the teleparallel gravity, instead of geodesics, there are force
equations, which are analogous to the Lorentz force equation of
electrodynamics \cite{trans} thus bridging the two theories upto
certain extents.

The paper is organized as follows. In section II, the basic
equations for the $f(T)$ model is discussed and the statefinder
parameters are calculated. In section III, a model of $f(T)$
gravity is considered, and the statefinder parameters are
evaluated. A detailed physical analysis is done in section IV.
Finally we end with some concluding remarks in section V.

\section{$f(T)$ gravity and dynamical equations}

$f(T)$ gravity developed as an alternative theory for GR, defined
on the Weitzenbock manifold, working only with torsion with no
curvature. A manifold is divided into two separate parts connected
with each other. One part having a Riemannian structure with a
definite metric described on it and another part having a
non-Riemannian structure with torsion or non-metricity. The part
having zero Riemannian tensor but having non zero torsion
represents a Weitzenbock spacetime. If $f(T) = T$ , this theory
reduces to teleparallel gravity \cite{hayashi, hehl}. It is seen
that with linear $T$ , this model has many common features like GR
and satisfies some standard tests of the GR in solar system
\cite{hayashi}. A general model of $f(T)$ was proposed only a few
years back \cite{f(T)}. This model has many important features.
Birkhoff's theorem has already been studied in this gravity
\cite{birkhoff}. The authors in \cite{zheng} investigated
perturbation in $f(T)$ and found that the perturbation in $f(T)$
gravity grows slower than that in Einstein GR. Bamba et al
\cite{bamba} studied the evolution of equation of state parameter
and phantom crossing in $f(T)$ model. Emergent Universes in
chameleon $f(T)$ model is investigated in \cite{chat}. Also it is
found that the dark matter problem can be addressed and resolved
in $f(T)$ gravity \cite{dm}. In this paper we intend to perform a
complete analysis of the statefinder parameters in various $f(T)$
models.

A suitable form of  action for $f(T)$ gravity in Weitzenbock
spacetime is  \cite{f(T)}
\begin{eqnarray}\nonumber
\mathcal{S}=\frac{1}{2\kappa^2}\int d^4x
e\left(T+f(T)+\mathcal{L}_m\right).
\end{eqnarray}
Here $e=det(e^{i}_{\mu})=\sqrt{-g}$, $\kappa^2=8\pi G$ and
$e^{i}_{\mu}$ is the tetrad (vierbein) basis. The dynamical
quantity of the model is the vierbein $e^{i}_{\mu}$ and $L_m$ is
the matter Lagrangian. The Friedmann equation in this form of the
$f(T)$ model \cite{f(T)} is given by,
\begin{equation}
H^2=\frac{1}{1+2f_T}\Big( \frac{\kappa^2}{3}\rho-\frac{f}{6}
\Big),\label{eq1}
\end{equation}
where $\rho=\rho_m$, while $\rho_m$ represents the energy densities
of matter. Here $f_{T}=\frac{df}{dT}$ and $f$ is a function of $T$.

Another FRW equation is
\begin{equation}
\dot
H=-\frac{\kappa^2}{2}\Big(\frac{\rho+p}{1+f_T+2Tf_{TT}}\Big).\label{eq2}
\end{equation}
Here $f_{TT}=\frac{d^{2}f}{dT^{2}}$. It is useful to rewrite the
equations (\ref{eq1}) and (\ref{eq2}) in the following suitable form
\begin{eqnarray}
H^2&=&\frac{\kappa^2}{3}\left(\rho_m+\rho_T\right),\label{eq3}\\
\dot
H&=&-\frac{\kappa^2}{2}\left(\rho_m+\rho_T+p_T\right),\label{eq4}
\end{eqnarray}
where
\begin{eqnarray}
\rho_T&=&\frac{1}{\kappa^2}\left(-\frac{f}{2}+Tf_T\right),\label{eq5}\\
p_T&=&\frac{1}{\kappa^2}\left[\frac{f}{2}+\left(2\dot
H-T\right)f_T+4T \dot Hf_{TT}\right].\label{eq6}
\end{eqnarray}\label{eq7}
The dimensionless density parameters are defined by
\begin{eqnarray}
\Omega_m+\Omega_T=1,\ \ \Omega_m=\frac{\kappa^2\rho_m}{3H^2},\ \
 \Omega_T=\frac{\kappa^2\rho_T}{3H^2}.
\end{eqnarray}
Easily we can write the following expression for deceleration
parameter $q$ and equation of state (EoS) parameter $w$,
\begin{eqnarray}
q=\frac{1}{2}\left(1+3w\Omega_T\right),\ \
~~~~w=\frac{p_T}{\rho_T}=-1+4\dot H\Phi,\ \ \Phi\equiv\frac{f_T+2T
f_{TT}}{-f+2T f_T}.\label{q}
\end{eqnarray}
Further, the evolutionary equation for $\Omega_T$ reads
\begin{eqnarray}
\dot \Omega_T=3H T
\left(\frac{f_T}{T}-\frac{f}{T^2}-2f_{TT}\right)\left(1+w
\Omega_T\right).\label{omegadot}
\end{eqnarray}
The state finder parameter $r$ can also be written as
\begin{eqnarray}
r=2q^2+q-\frac{\dot q}{H}.\label{rs}
\end{eqnarray}
Using equations (\ref{q}) and (\ref{omegadot}) in (\ref{rs}) we
obtain the statefinder parameters for $f(T)$ cosmology:
\begin{eqnarray}
r&=&\frac{1}{2}\left[\Big(1+3\left(-1+4\dot
H\Phi\right)\Omega_T\Big)^2+\Big(1+3\left(-1+4\dot
H\Phi\right)\Omega_T\Big)\right]\nonumber\\&& -\frac{3}{2H}\Big[\dot
w \Omega_T+3w H \left(1+w
\Omega_T\right)\left(\frac{f_T}{T}-\frac{f}{T^2}-2f_{TT}\right)\Big],
\end{eqnarray}
\begin{equation}
s=\frac{2(r-1)}{9(-1+4\dot H\Phi) \Omega_T}.
\end{equation}
where
\begin{equation}
\dot {w}=4\Big(\ddot H \Phi+\dot H \dot T \frac{d\Phi}{dT}\Big).
\end{equation}

\begin{figure*}[thbp]
\begin{tabular}{rl}
\includegraphics[width=7.5cm]{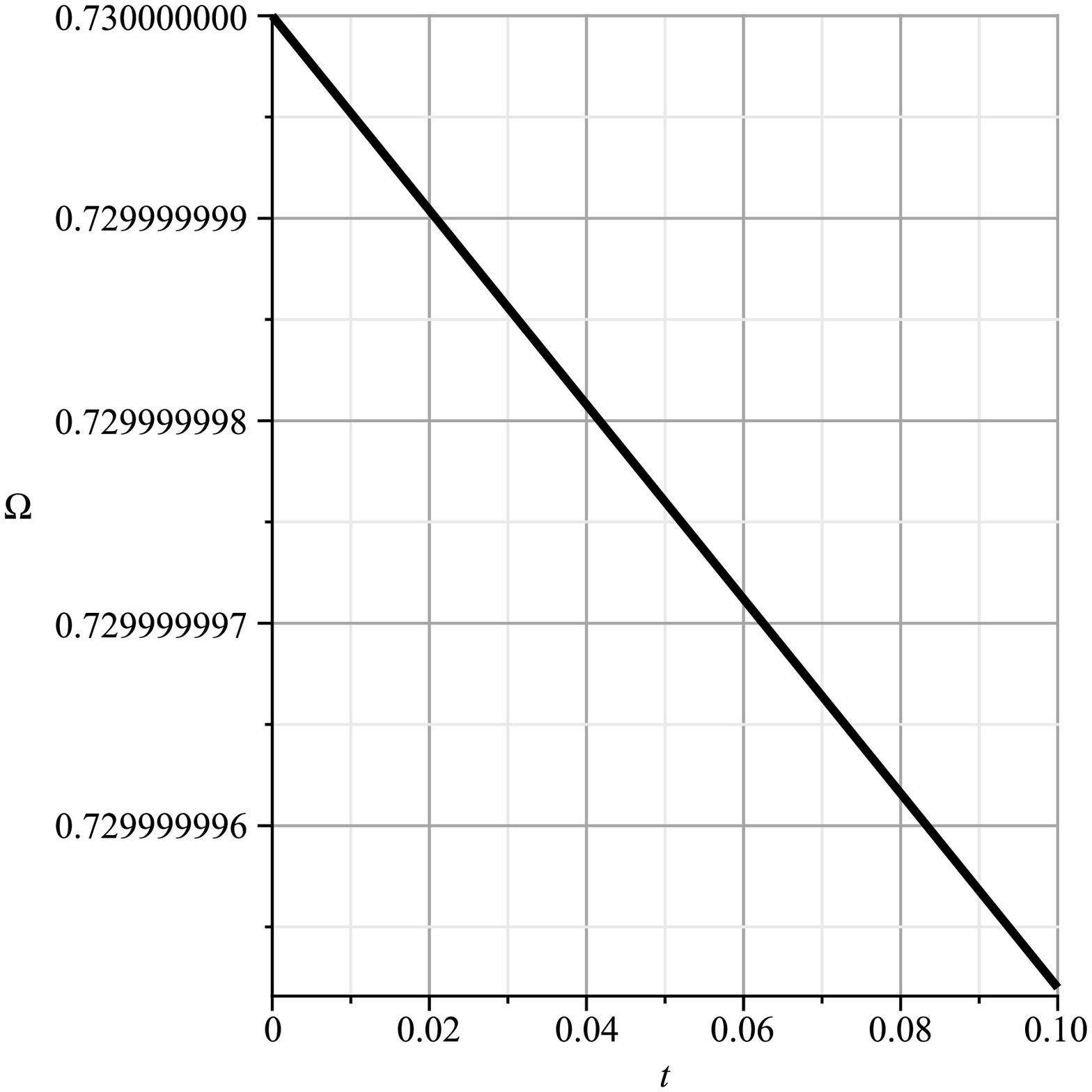}&
\includegraphics[width=7.5cm]{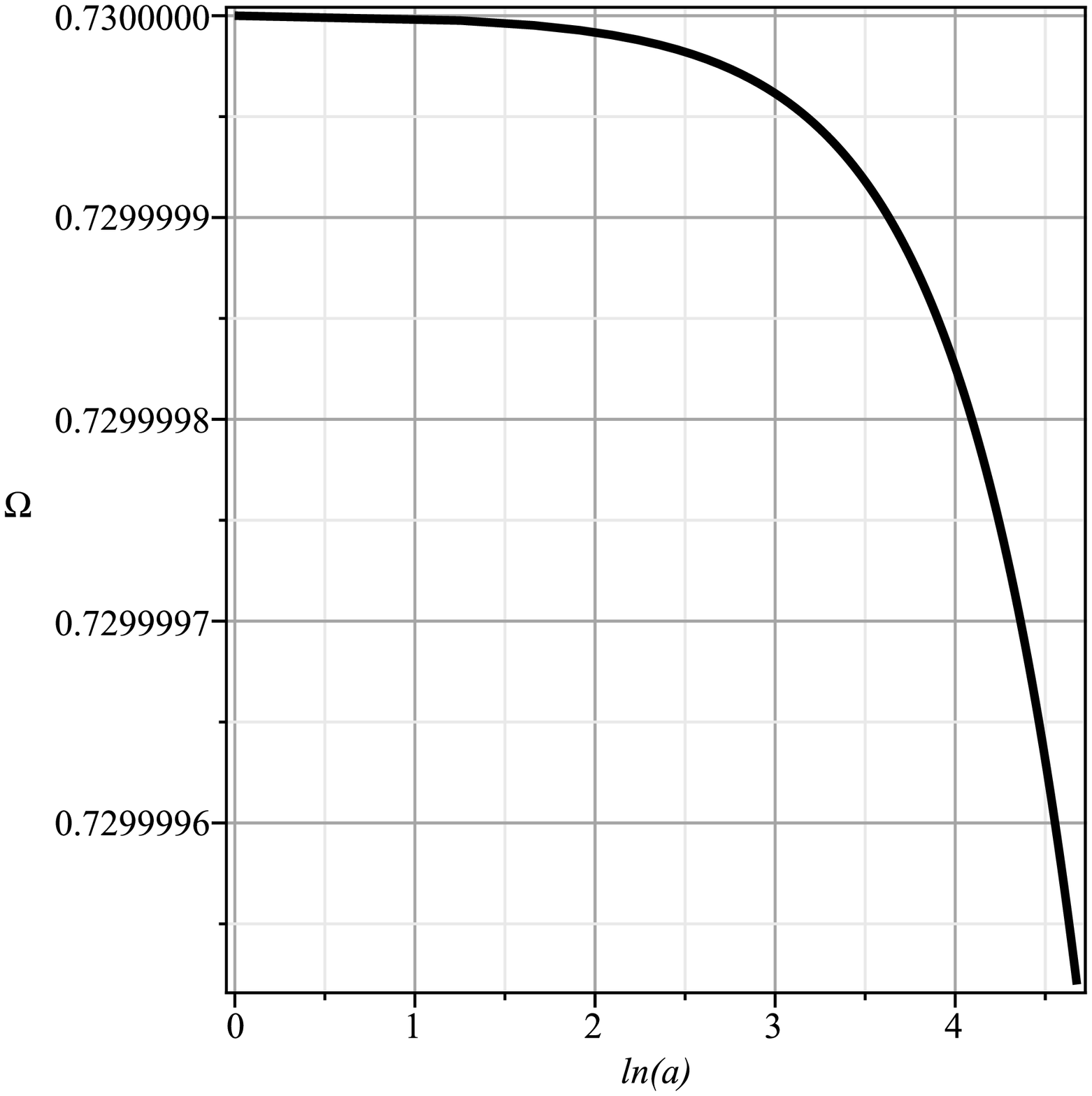} \\
\includegraphics[width=7cm]{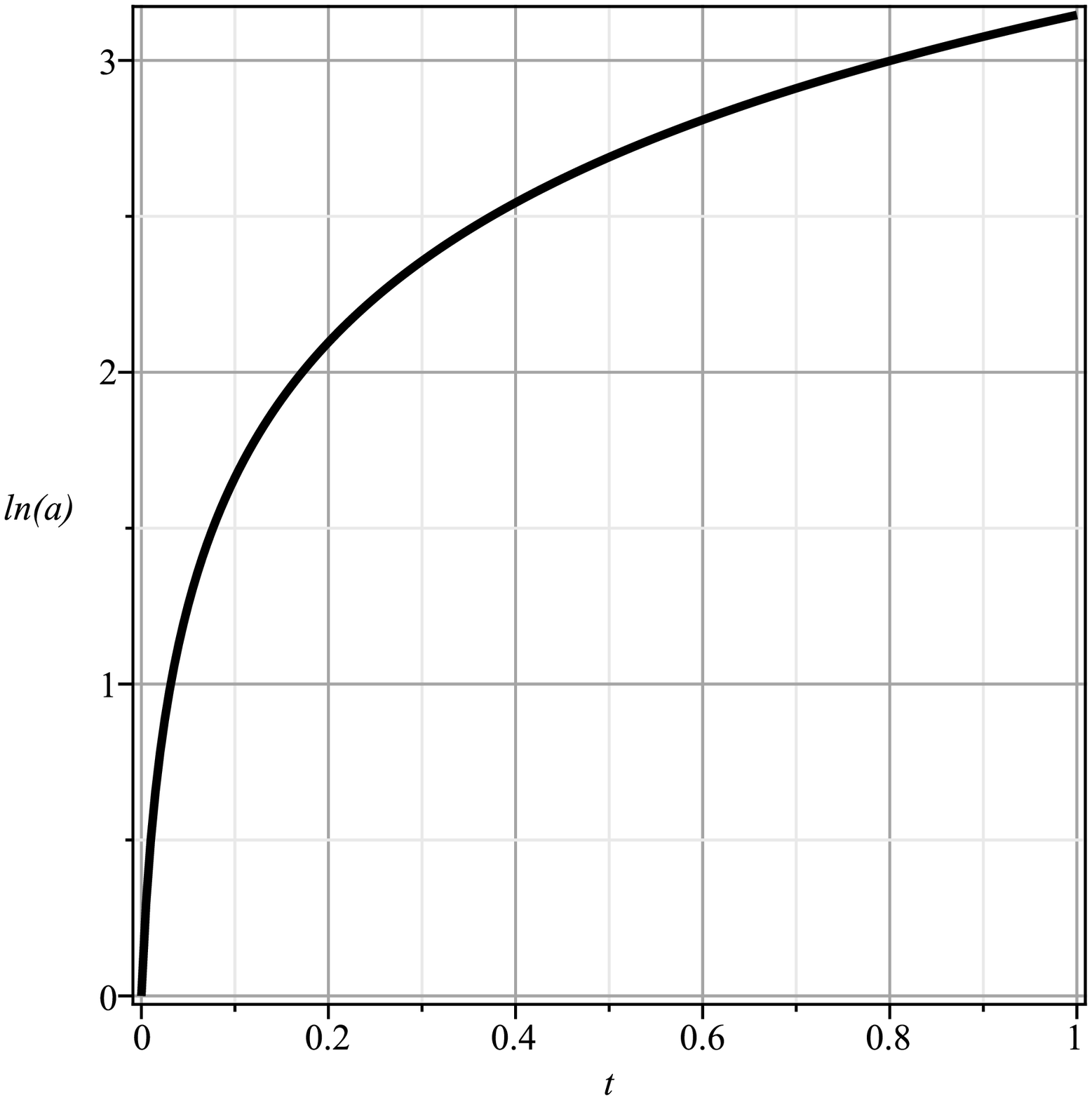}&
\includegraphics[width=7cm]{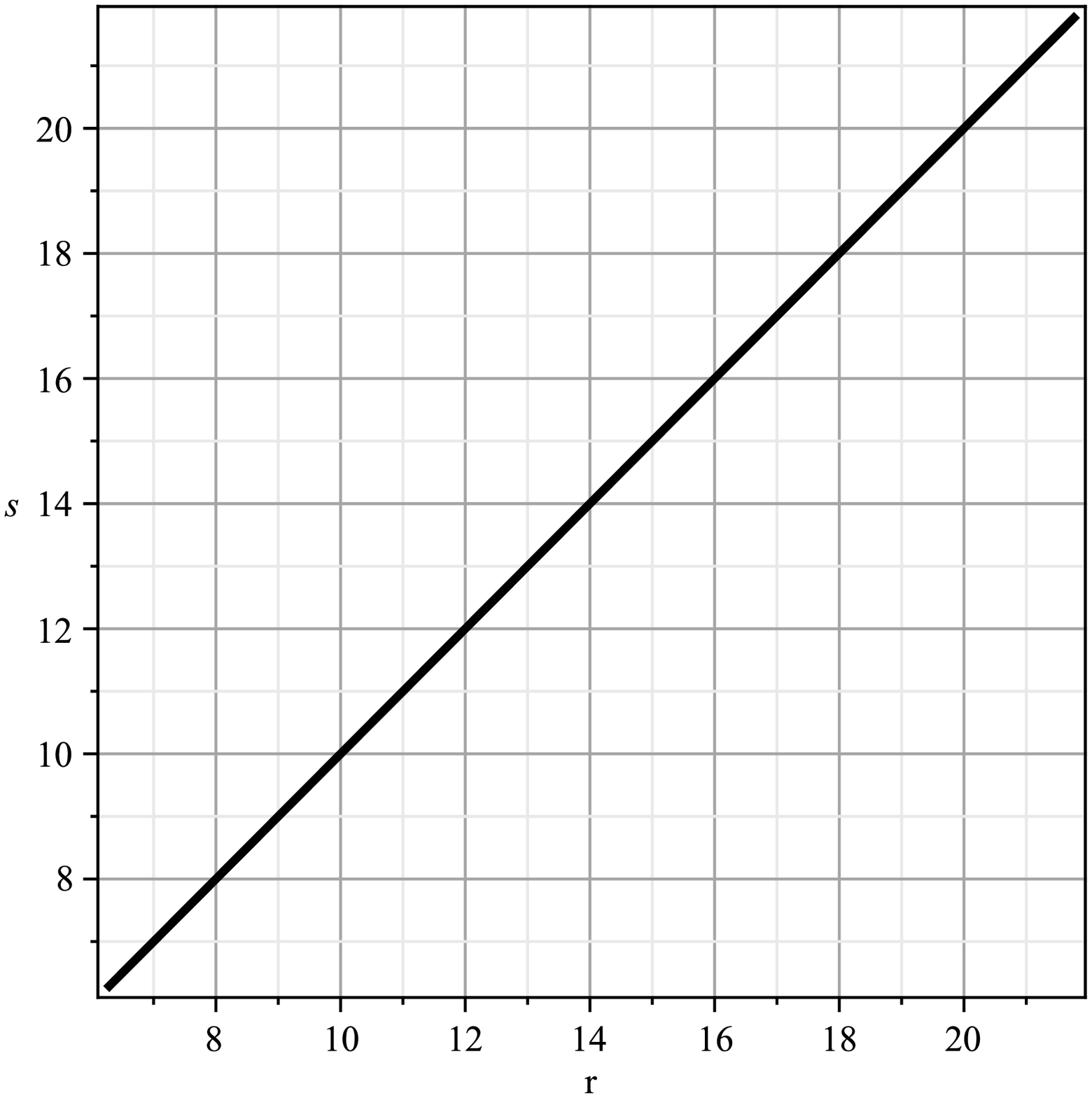} \\
\includegraphics[width=7cm]{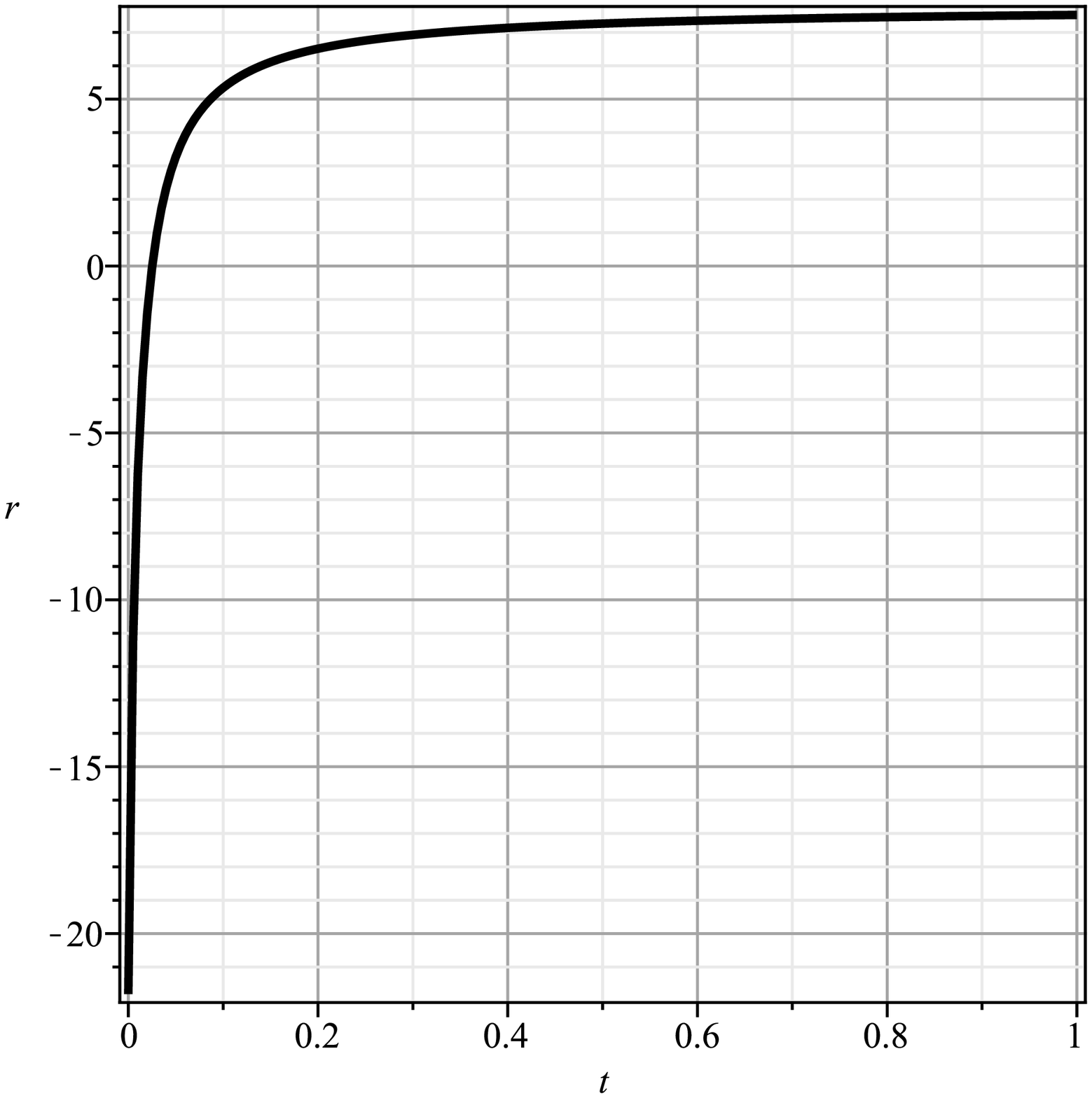}&
\includegraphics[width=7cm]{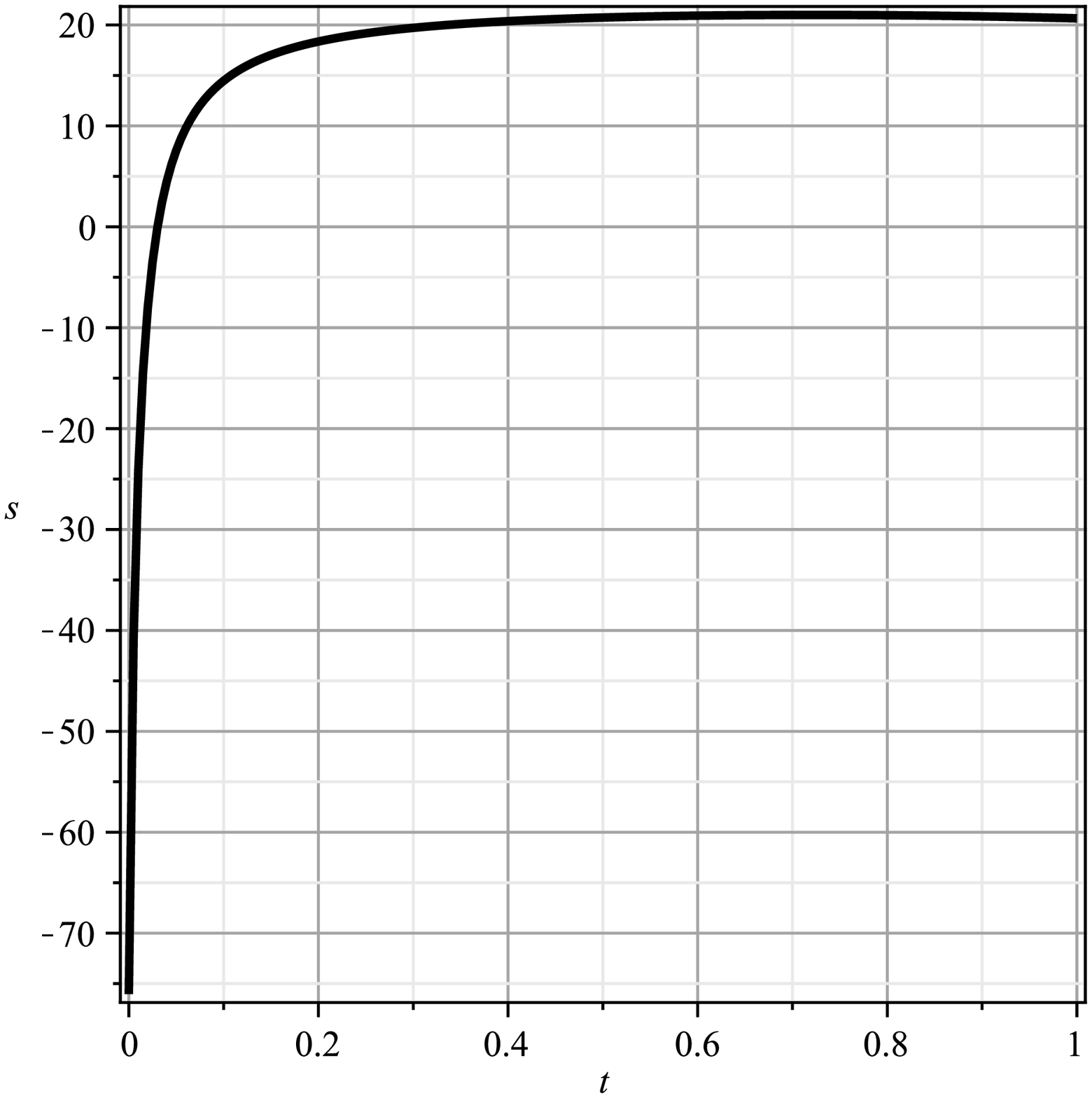} \\
\end{tabular}
\caption{ (\textit{Top Left}) Cosmic energy density parameter
$\Omega$ against time. (\textit{Top Right}) Cosmic energy density
parameter $\Omega$ against logarithmic scale factor. (\textit{Middle
Left}) Evolution of scale factor with time. (\textit{Middle Right})
Statefinder parameters $\{r,s\}$ against each other. (\textit{Bottom
Left}) Parameter $r$ vs time. (\textit{Bottom Right}) Parameter $s$
vs time. While plotting these figures, we assumed
$\Omega_{0m}\simeq0.2$, $C_2=0$, $\alpha=0.2$,
$\Omega_{d0}\simeq0.73$, $H_0\simeq74$. }
\end{figure*}

\begin{figure*}[thbp]
\begin{tabular}{rl}
\includegraphics[width=7.5cm]{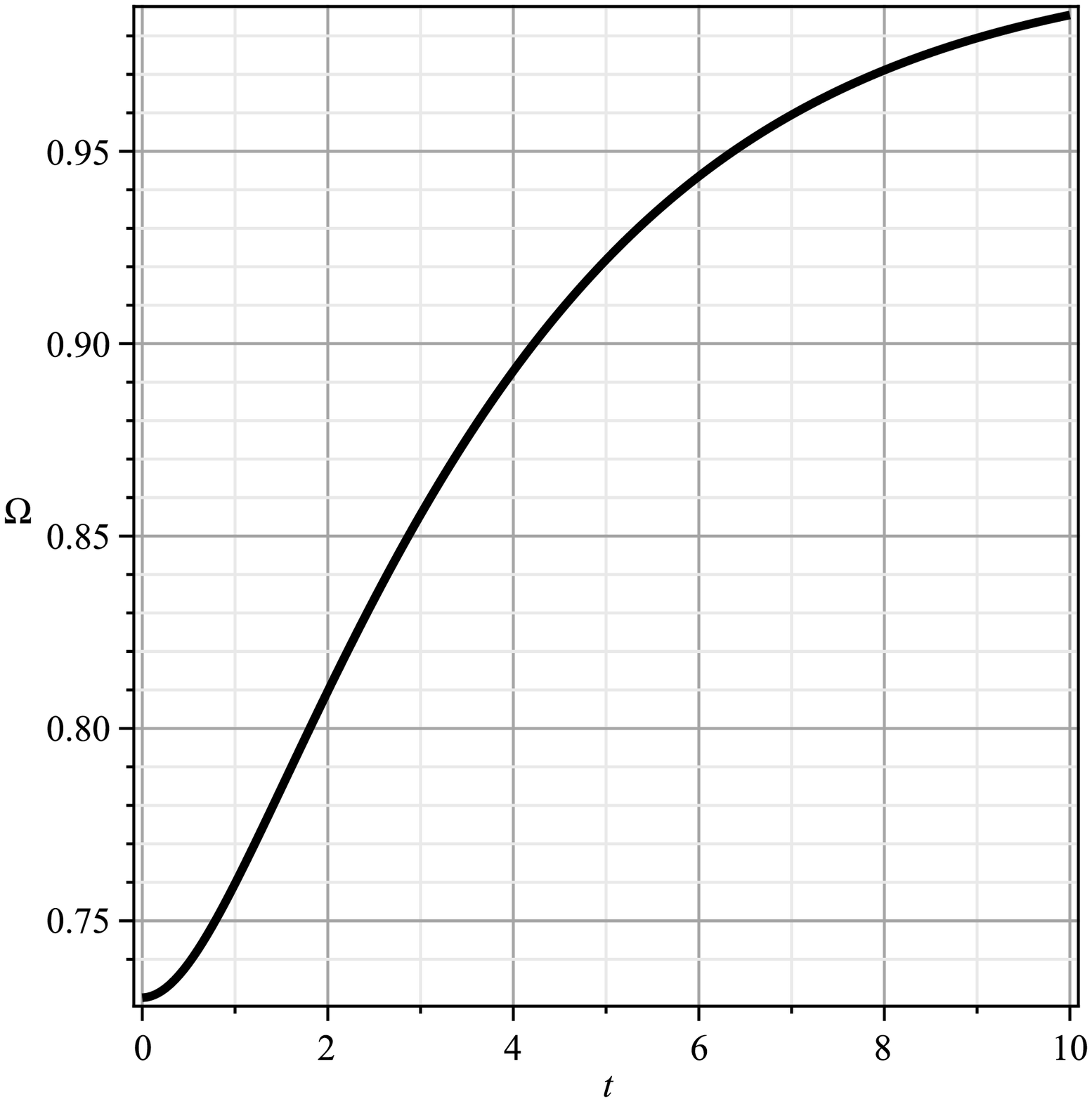}&
\includegraphics[width=7.5cm]{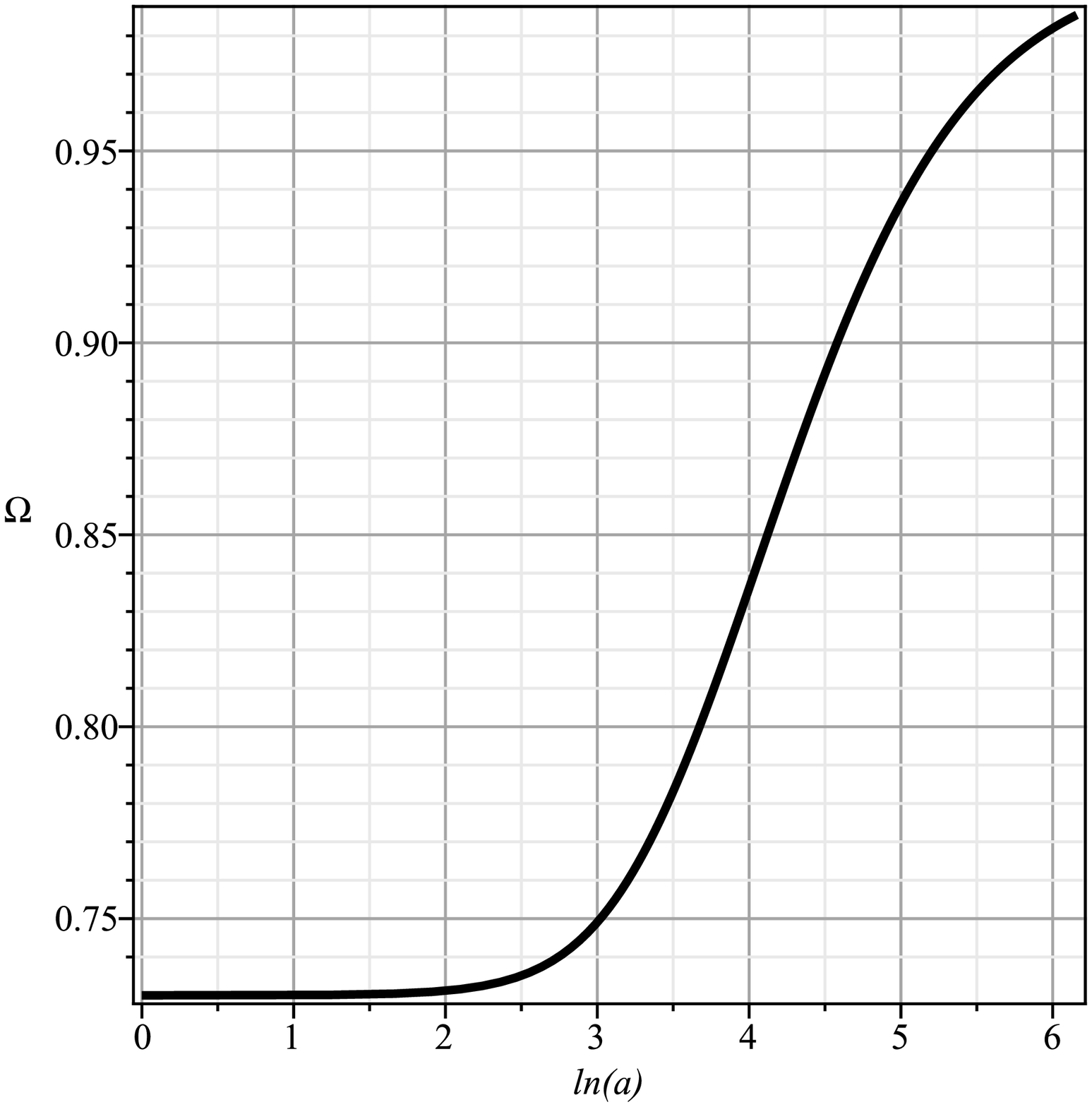} \\
\end{tabular}
\caption{ (\textit{Left}) Cosmic energy density parameter $\Omega$
against time for teleparallel gravity. (\textit{Right}) Cosmic
energy density parameter $\Omega$ against logarithmic scale factor
for teleparallel gravity. Here we put $C_1=0$ and $C_2=0.1$ and used
the same initial conditions as Figure-1.}
\end{figure*}

\section{Statefinder parameters for a particular $f(T)$}

In order to avoid analytic and computation problems, we choose a
suitable expression for $f(T)$  which contains a constant, linear
and a non-linear form of torsion, specifically \cite{attractor}
\begin{equation}\label{eqm1}
f(T)=2C_1 \sqrt{-T} +\alpha T+C_2,
\end{equation}
where $\alpha$, $C_1$ and $C_2$ are arbitrary constants. Note that
choosing $C_1=0$ in (\ref{eqm1}) leads to teleparallel gravity.

In this model the combination of the first and the third term
corresponds to the EoS of the cosmological constant in the framework
of $f(T)$ gravity \cite{mirza}. In this way by shuffling various
terms or by the introduction of new terms, cosmologists all over the
world have succeeded in establishing different models. In fact many
of them have been reconstructed from various dark energy models. The
model in equation (\ref{eqm1}) that we are currently dealing with
may have been inspired from the proposed model of Veneziano ghost
\cite{kk}.

As a matter of fact this model was preferentially chosen over other
alternatives, purely because of its simplicity as far as numerical
computation is concerned. Another advantage of this model is that
the obtained results are easier to compare or differentiate from the
corresponding results in GR. In order to facilitate this, the linear
middle term is included in the model. In connection with this model,
it is worth stating that Capozziello et al \cite{capo123} made an
attempt to investigate the cosmography of $f(T)$ cosmology by using
data from BAO, Supernovae Ia and WMAP. The analysis performed by
Capozziello and his colleagues unveiled the fact that by choosing
$C_{2} = 0$,~ $\alpha =\Omega_{m0}$ and $C_{1} =
\sqrt{6}H_{0}\left(\Omega_{m0}-1\right)$, it is possible to estimate
the parameters of our proposed $f(T)$ model as functions of Hubble
parameter $H_{0}$, the cosmographic parameters and the value of
matter density parameter.

Using this model we have the following expressions for the
statefinder parameters,

\begin{eqnarray}
r&=&\frac {1}{72 \left( H \left( t \right)  \right) ^{5}}\left[144\,
 \left( {\frac {d}{dt}}H \left( t\right)  \right) ^{2}H \left( t \right)
  \Omega_{{{\it m0}}}+144\,H\left( t \right) {\Omega_{{{\it m0}}}}^{2}
   \left( {\frac {d}{dt}}H\left( t \right)  \right) ^{2}+432\, \left( H
   \left( t \right)\right) ^{3}{\Omega_{{{\it m0}}}}^{2}
   {\frac {d}{dt}}H \left( t\right)\right.\nonumber\\&&
\left.+216\, \left( H \left( t \right)  \right) ^{3}\Omega_{{{\it
m0}}}{\frac {d}{dt}}H \left( t \right) -72\, \left(H \left( t
\right)  \right) ^{2} \left( {\frac {d^{2}}{d{t}^{2}}}H \left(
t\right)  \right) \Omega_{{{\it m0}}}+72\, \left( H \left(
t\right)\right) ^{5}+324\, \left( H \left( t \right)
\right)^{5}\Omega_{{{ \it m0}}}\right.\nonumber\\&& \left.+324\,
\left( H \left( t \right) \right)^{5}{\Omega_{{{\it
m0}}}}^{2}\right],
\end{eqnarray}

\begin{eqnarray}
s&=&\frac{1}{18\left( H \left( t \right)
 \right) ^{3} \left( 6\, \left( H \left( t \right)  \right) ^{2}+4\,{
\frac {d}{dt}}H \left( t \right)\right)\,} \left[48\left( {\frac
{d}{dt}}H \left( t \right) \right) ^ {2}H \left( t \right) +48\,
\left( {\frac {d}{dt}}H \left( t \right)
 \right) ^{2}H \left( t \right) \Omega_{{{\it m0}}}\right.\nonumber\\&&
\left. +144\, \left( H\left( t \right)  \right) ^{3}\Omega_{{{\it
m0}}}{\frac {d}{dt}}H
 \left( t \right) +72\, \left( H \left( t \right)  \right) ^{3}{\frac
{d}{dt}}H \left( t \right)-24\, \left( {\frac {d^{2}}{d{t}^{2}}}H
 \left( t \right)  \right)  \left( H \left( t \right)  \right) ^{2}+
108\, \left( H \left( t \right)  \right) ^{5}\right. \nonumber\\&&
\left.+108\, \left( H \left( t \right)  \right) ^{5}\Omega_{{{\it
m0}}}\right],
\end{eqnarray}
where $\Omega_{{{\it m0}}}$ is the present dimensionless fractional
matter density, such that $\Omega_{{{\it m0}}}=\frac{\rho_{{{\it
m0}}}}{3H_{0}^{2}}$. Here $\rho_{{{\it m0}}}$ is the present matter
density and $H_{0}$ is the present Hubble parameter. In accordance
with the current observational data the value of $\Omega_{{{\it
m0}}}$ is almost equal to $0.27$. It is interesting to note that for
present time the $\{r,s\}$ parameters take the form
\begin{eqnarray}
r&=&1+4.5\Omega_{\it m0}(1+\Omega_{\it m0}),\\ s&=&1+\Omega_{\it
m0}.
\end{eqnarray}

\section{Numerical results and physical aspects}

In this section, we will solve the time evolutionary equations
(\ref{eq4}) and (\ref{omegadot}) by numerical algorithms. We adopt
the initial conditions as $\Omega_{0m}=0.2$, $C_2=0$, $\alpha=0.2$,
$\Omega_{d0}=0.73$, $H_0=74$. It is observed that these are regular
solutions and the generic form of the solutions is independent from
the initial conditions imposed on the functions. In figure-1, the
top left and right figures represent the evolution of density
parameter $\Omega_T$ against time and logarithmic scale factor. In
either figures, it is easily seen the energy density parameter
decreases and vanish in the far future. This shows that if dark
energy is represented by the torsion of space, it will decay in far
future. Moreover the middle left figure in the same panel shows that
the slope of logarithmic scale factor is concave downwards, i.e. the
model predicts deceleration on account of decaying dark energy.
Middle right figure shows the trajectory of the chosen $f(T)$ in the
$\{r,s\}$ plane. The bottom left and right figures give the
evolutionary trajectories of $r$ and $s$ independently. It is
interesting to note that trajectories of both parameters are
essentially identical except the rescaling of the vertical
coordinate. In figure-2, we plot figures for the teleparallel
gravity by choosing $C_1=0,$ and adopting the same initial
conditions. Contrary to the previous case, the teleparallel gravity
predicts the increase in the dark energy density i.e.
$\Omega_T\rightarrow1.$ In other words, the dark energy component
due to torsion will overcome on all the remaining energy content of
the Universe.

\section{Conclusion}

We have performed a complete analysis of statefinder parameters in
$f(T)$ gravity, which is a mere modification of the teleparallel
gravity once proposed by Einstein. We considered a model containing
linear and non-linear terms of $T$) have been considered and a
complete diagnosis of the statefinder parameters have been carried
out. Plots have been generated for the statefinder parameters $r$
and $s$ with respect to the Hubble parameter $H$, for different
values of model parameters.  This gives us a clear notion about the
nature of the modified gravity theory $f(T)$ and its governing
dynamics. Specifically we found that our model $f(T)=2C_1 \sqrt{-T}
+\alpha T+C_2,$ predicts of decay of dark energy in the far future
while its special case namely teleparallel gravity predicts that
dark energy will overcome over all the energy content of the
Universe. We hope that by comparing the statefinder analysis
illustrated in this work with the future observational data, $f(T)$
models can be distinguished from the $\Lambda$CDM model.

\subsection*{\bf Acknowledgments}

We would like to thank anonymous referee for giving useful comments
to improve this paper.


\begin{thebibliography}{99}
\bibitem{c1}
S. Perlmutter  et al., Astrophys. J. 517, 565 (1999); D. N. Spergel
et al. Astron. J. Suppl 148, 175 (2003).

\bibitem{c2}
C.L. Bennett et al., Astrophys. J. Suppl. 148, 1 (2003).
\bibitem{c3}
M. Tegmark et al., Phys. Rev. D 69, 103501 (2004).
\bibitem{c4}
S.W. Allen et al., Mon. Not. Roy. Astron. Soc. 353, 457 (2004).


\bibitem{Riess1} A.G.~Riess  et al., Astron.\ J.\  116, 1009
(1998).

\bibitem{bamba1} K. Bamba, S. Capozziello, S. Nojiri, S.D. Odintsov,
Astrophysics and Space Science, DOI: 10.1007/s10509-012-1181-8
[arXiv:1205.3421v3 [gr-qc]]

\bibitem{c7}
V. Sahni, A. Starobinsky, Int. J. Mod. Phy. D 9, 373 (2000); P.J.
Peebles, B. Ratra, Rev. Mod. Phys. 75, 559 (2003);



\bibitem{quint}
B.~Ratra, P.J.E.~Peebles, Phys.\ Rev.\ D 37, 3406 (1988);
C.~Wetterich, Nucl.\ Phys.\ B 302, 668 (1988); A.~R.~Liddle,
R.J.~Scherrer, Phys.\ Rev.\ D 59, 023509 (1999).

\bibitem{chaplygin} A. Kamenshchik, U. Moschella, V. Pasquier,
Phys. Lett. B 511, 265 (2001); M. Jamil, M.A. Rashid, Eur. Phys. J.
C 58, 111 (2008); M. Jamil, M.A. Rashid, Eur. Phys. J. C 60, 141
(2009); M. Jamil, Int. J. Theor. Phys. 49, 62 (2010); M. Jamil, Int.
J. Theor. Phys. 49, 144 (2010); M.U. Farooq, M.A. Rashid, M. Jamil,
Int. J. Theor. Phys. 49, 2334 (2010).


\bibitem{phant} R. R. Caldwell, Phys.
Lett. B 545, 23 (2002); R.R.~Caldwell, M.~Kamionkowski,
N.N.~Weinberg, Phys. Rev. Lett. 91, 071301 (2003); M. Jamil, S. Ali,
D. Momeni, R. Myrzakulov, Eur. Phys. J. C 72, 1998 (2012); M. Jamil,
D. Momeni, M. Raza, R. Myrzakulov, Eur. Phys. J. C 72, 1999 (2012).

\bibitem{f} M. Jamil, Y. Myrzakulov, O. Razina, R. Myrzakulov, Astrophys. Space Sci.
336, 315 (2011); M. Jamil, D. Momeni, N. S. Serikbayev, R.
Myrzakulov, Astrophys. Space Sci. 339, 37 (2012).


\bibitem{obs} H. Wei, S. N. Zhang, Phys. Lett. B 654, 139 (2007);
 M-L Tong, Y Zhang, Z-W Fu, Class. Quant. Grav. 28, 055006 (2011);
Y-H Li, J-Z Ma, J-L. Cui, Z. Wang, X. Zhang, Sci. China Phys. Mech.
Astron. 54, 1367 (2011); H. Wei, Phys. Lett. B 691, 173 (2010); S.
M. R. Micheletti, JCAP 1005, 009 (2010); C. Feng, B. Wang, E.
Abdalla, R-K. Su, Phys. Lett. B 665, 111 (2008); L. Amendola, C.
Quercellini, D. T-Valentini, A. Pasqui, Astrophys.J. 583, L53
(2003).

\bibitem{cr} G. Olivares, F. A-Barandela, D. Pavon, Phys. Rev. D 77, 063513 (2008).

\bibitem{pr} M Jamil, A Sheykhi, M. U Farooq, Int. J. Mod. Phys.
D 19, 1831 (2010); K. Karami, A. Sheykhi, M. Jamil, Z. Azarmi, M. M.
Soltanzadeh, Gen. Relativ. Grav. 43, 27 (2011); M. Jamil, E. N.
Saridakis, JCAP 1007, 028 (2010); P. Rudra, U. Debnath, R. Biswas,
Astrophys Space Sci (2012) 339:53–64; R. Chowdhury, P. Rudra,
[arXiv:1204.3531 [gr-qc]];  M. Jamil, M.A. Rashid, Eur. Phys. J. C
56, 429 (2008); M. Jamil, F. Rahaman, Eur. Phys. J. C 64, 97 (2009).


\bibitem{Sahni1} V. Sahni, T.D. Saini, A.A. Starobinsky, U. Alam,
 JETP Lett. 77, 201 (2003).


\bibitem{jamil} M.R. Setare, M. Jamil, Gen. Relativ. Gravit. 43, 293
(2011);  M. Jamil, M. Raza, U. Debnath, Astrophys. Space Sci. 337,
799 (2012); S. Chakraborty, U. Debnath, M. Jamil, Canadian J. Phys.
90, 365 (2012); M. Jamil, I. Hussain, D. Momeni, Eur. Phys. J. Plus
126, 80 (2011); U. Debnath, M. Jamil, Astrophys. Space Sci. 335, 545
(2011); M. Jamil, Int. J. Theor. Phys. 49, 2829 (2010); M. Jamil, U.
Debnath, Int. J. Theor. Phys. 50, 1602 (2011).

\bibitem{pano} G. Panotopoulos, Nucl.Phys.B. 796, 66-76 (2008)


\bibitem{sergei2}
S. Nojiri, S. D. Odintsov,  Phys. Rept. 505, 59 (2011); S.
Capozziello, M. De Laurentis, arXiv:1108.6266v2 [gr-qc]; T. Clifton,
P. G. Ferreira, A. Padilla, C. Skordis, arXiv:1106.2476v2
[astro-ph.CO]

\bibitem{lqc} C. Rovelli, liv. Rev. Rel.1, 1 (1998);
 A. Ashtekar, J. Lewandowski, Class. Quantum. Grav.
 21, R53 (2004); M. Jamil, D. Momeni, M.A. Rashid,
 Eur. Phys. J. C 71, 1711 (2011); M. Jamil, U. Debnath, Astrophys. Space
Sci. 333, 3 (2011); D. Dwivedee, B. Nayak, M. Jamil, L.P. Singh,
arXiv:1110.6350v2 [gr-qc].

\bibitem{brane} Maartens, R., Phys. Rev. D 62, 084023
(2000); R. Maartens, Living Rev. Relativity 7, 7 (2004).

\bibitem{fr} R. Kerner, Gen. Relativ. Gravit. 14, 453 (1982);
 G. Allemandi, A. Borowiec, M. Francaviglia, Phys. Rev. D 70, 103503
(2004); S.M. Carroll, A.D. Felice, V.Duvvuri, D.A. Easson, M.
Trodden, M.S. Turner, Phys. Rev. D 71, 063513 (2005); T.P. Sotiriou,
V. Faraoni, gr-qc/0805.1726.

\bibitem{ft} K. K. Yerzhanov et al, arXiv:1006.3879v1 [gr-qc];
 R. Myrzakulov (2010), arXiv:1006.1120v1 [gr-qc]; E. V.
Linder, Phys. Rev. D 81 127301 (2010).

\bibitem{miao} M. Li, R-X Miao, Y-G Miao, JHEP 1107, 108 (2011).

\bibitem{miaoli}
R. Miao, M. Li, Y. Miao, JCAP 11, 033 (2011).

\bibitem{sotirio}
B. Li, T. P. Sotiriou, J. D. Barrow, Phys. Rev. D 83, 064035
(2011).


\bibitem{trans} V. C. de Andrade, J. G. Pereira, Phys. Rev D 56, 4689
(1997); F. W. Hehl, J. D. McCrea, E. W. Mielke, Y. Ne'eman, Phys.
Rep. 258, 1 (1995).

\bibitem{hayashi}
    K. Hayashi, T. Shirafuji, Phys. Rev. D 19, 3524 (1979);
    K. Hayashi, T. Shirafuji, Phys. Rev. D 24, 3312 (1981).


\bibitem{hehl}
F.  Hehl, P. von der Heyde, G.  Kerlick, Rev. Mod. Phys. 48, 393
(1976).

\bibitem{f(T)}
R.Ferraro, F.Fiorini, Phys. Rev. D 75, 084031 (2007); R. Ferraro, F.
Fiorini, Phys. Rev. D 78, 124019 (2008).

\bibitem{birkhoff}
X. Meng, Y. Wang, Eur. Phys. J. C 71 (2011) 1755.

\bibitem{zheng} R. Zheng, Q-G. Huang, JCAP 1103, 002 (2011).

\bibitem{bamba} K. Bamba, C-Q Geng, C-C Lee, L-W Luo, JCAP 1101, 021
(2011).

\bibitem{chat} S. Chattopadhyay, U. Debnath, Int. J. Mod. Phys. D 20,
1135 (2011).

\bibitem{dm} M. Jamil, D. Momeni, R. Myrzakulov, Eur. Phys. J. C 72,
2122 (2012).


\bibitem{attractor}
 M. Jamil, D. Momeni, R. Myrzakulov, Eur. Phys. J. C 72, 1959 (2012).

\bibitem{mirza} R. Myrzakulov, Eur. Phys. J. C 71, 1752 (2011).

\bibitem{kk}
K. Karami, A. Abdolmaleki, arXiv:1202.2278.

\bibitem{capo123} S. Capozziello, V.F. Cardone, H. Farajollahi,
 A. Ravanpak, Phys. Rev. D 84, 043527 (2011).



\end{thebibliography}
\end{document}